\title{CSCW 2018: Submitted for Second Round}
\documentclass[format=acmsmall, review=false, screen=true]{acmart}

\usepackage{booktabs} 

\usepackage[ruled]{algorithm2e} 

\SetAlFnt{\small}
\SetAlCapFnt{\small}
\SetAlCapNameFnt{\small}
\SetAlCapHSkip{0pt}
\IncMargin{-\parindent}

\acmJournal{PACMHCI}
\acmVolume{2}
\acmNumber{CSCW}
\acmArticle{56}
\acmYear{2018}
\acmMonth{11}
\copyrightyear{2018}

\setcopyright{acmlicensed}

\acmDOI{10.1145/3274325}

\begin{document}
\title[]{Social Photo-elicitation: The Use of Communal Production of Meaning to Hear a Vulnerable Population}
\author{Aakash Gautam}
\affiliation{%
  \institution{Virginia Tech}
  \city{Blacksburg}
  \state{VA}
  \postcode{24060}
  \country{USA} }
\email{aakashg@vt.edu}
\author{Chandani Shrestha}
\affiliation{%
  \institution{Virginia Tech}
  \city{Blacksburg}
  \state{VA}
  \postcode{24060}
  \country{USA} }
\email{chandani@vt.edu}
\author{Steve Harrison}
\affiliation{%
  \institution{Virginia Tech}
  \city{Blacksburg}
  \state{VA}
  \postcode{24060}
  \country{USA} }
\email{srh@cs.vt.edu}
\author{Deborah Tatar}
\affiliation{%
  \institution{Virginia Tech}
  \city{Blacksburg}
  \state{VA}
  \postcode{24060}
  \country{USA} }
\email{dtatar@cs.vt.edu}
\begin{abstract}

We report on an initial ethnographic exploration of the situation of sex-trafficking survivors in Nepal. In the course of studying trafficking survivors in a protected-living situation created by a non-governmental organization in Nepal, we adapted photo-elicitation to hear the voices of the survivors by making the technique more communal.  Bringing sociality to the forefront of the method reduced the pressure on survivors to assert voices as individuals, allowing them to speak. We make three contributions to research. First, we propose a communal form of photo-elicitation as a method to elicit values in sensitive settings. Second, we present the complex circumstances of the survivors as they undergo rehabilitation and move towards life with a ``new normal''. Third, our work adds to HCI and CSCW literature on understanding specific concerns of trafficking survivors and aims to inform designs that can support reintegration of survivors in society. The values that the survivors hold and their notion of future opportunities suggest possession of limited but important social capital in some domains that could be leveraged to aid reintegration. 

\end{abstract}

%
%



\maketitle

\renewcommand{\shortauthors}{Gautam et al.}

\section{Introduction}

Sex-trafficking is acute in Nepal. The open border between Nepal and India is one of the busiest human trafficking sites in the world with between 5,000 and 12,000 girls trafficked annually \cite{nhrc2016}. There is additional trafficking within Nepal. Young girls and women are most vulnerable due to pervasive gender inequity and discrimination promoted by the prevalence of extreme poverty. Moreover, the decade-long civil war, the 2015 earthquake, and the on-going political instability have increased vulnerability and thereby contributed to the problem \cite{crawford2008sex, kaufman2012let, earthquakeIncreaseTrafficking}.

Non-Governmental Organizations (NGOs) implement some measures to prevent trafficking, and rescue and reintegrate survivors. Typically, these organizations offer a protected living situation with some form of counseling and skill-based training. These programs are very important but cannot claim universal success.  They have been criticized for the design of prevention programs \cite{kaufman2011research}, lack of evaluation of the interventions  \cite{crawford2008sex,kaufman2011research}, and overall efficacy of rehabilitation \cite{joshi2001cheli}. 

These programs seek to do something, against the odds.  But they are limited by many factors.  In general, their ultimate goal is to help survivors reintegrate into society, most preferably into the family from which the survivor originated.  This can be a problem for many reasons including that many in Nepali society believe that the return of the survivor brings disgrace not just to the individual but to her family and even the entire community \cite{mahendra2001community}. Acceptance is difficult at best \cite{frederick2000fallen}.  Sometimes the survivors return to the same family that sold her the first time with conditions essentially unchanged \cite{hennink2004sex}. 

The goal of the work presented in this paper was to understand whether and how we could aid the survivors in the reintegration process. To that end, the first author conducted an \emph{initial ethnographic inquiry}, working to understand the current circumstances of survivors and their notions about the future as they experience life and training in the protected living situation.  The context we study is sensitive in four ways: (1) The participants are vulnerable in a number of ways: young, uneducated, impoverished and already abused; (2) participants are dependent on the organization that has rescued them; (3) researchers are dependent on the organization for access to the population; and, (4) the organization itself is subjected to complex political, economic and cultural forces. It is worth noting that in 2016, 60.7\% of the annual funds of the organization discussed in this paper were from donors. 

As part of the initial ethnographic observation, we were led to create a \emph{social} value-eliciting activity. This paper reports on the adaptation of the photo-elicitation method, our experience, and findings of these initial attempts to hear the voice of the survivors in the rehabilitation program. We heard about their perseverance, their desire to belong in society with dignity and respect, their mixed emotions about crafting work, their sense of sisterhood and their dreams of family-like connectedness. We also heard something of their pain from the past events, a fear of being stigmatized by the society which they aspire to rejoin, and a limited notion of possibilities for economic independence. We located their social capital \cite{lin2017building,putnam2000bowling,granovetter1977strength,dillahunt2014fostering} in two strains---reliance on one another and crafting work---that the survivors could leverage as they proceed in the reintegration process. Supporting those strains could define an enactable design space for us, but we also consider drawbacks that would constitute moral and ethical barriers to our design action.

\subsection{Reflexivity and Commitments}
The first and second authors were born and raised in Nepal. The fieldwork was conducted by the first author. The analysis presented in this work is shaped by our views of the patriarchal values and norms in Nepal as problematic, which oppresses women and forces them to accept harrowing experiences including trafficking and abuse. In this way, the work undertaken here falls under the rubric of Feminist HCI \cite{bardzell2010feminist}. However, it starts with considerations of power and capability that are more oppressive and less optimistic. Pervasive patriarchal values and poverty, and their subsequent impact on increasing gender inequality in Nepal have been discussed in research work (e.g. \cite{kaufman2012let,menger2015unveiling}) leading one research work to term women in Nepal as ``second class citizens'' \cite{cameron2017women}. 

Our commitment is to a vision of what ``better'' might be.  Importantly, this vision may not be shared by the participants.  There is some evidence of wide-spread endorsement by women of the idea that a husband has the right to beat his wife, with claims ranging from 30\% endorsement in the region around Kathmandu to almost 80\% in some rural regions \cite{yoshikawa2014acceptance,lamichhane2011women}. We can understand the circumstances and cultural history that lead to that but not accept it.

However, despite our differences from the population that we seek to serve, our intention is to listen to them, hear their values, understand their perception of future possibilities, and, if relevant, engage in the design of interventions to help support the survivors.  Furthermore, we seek to learn from both positive and negative examples of intervention.  

\section{Related Work}

\subsection{Vulnerable Communities and HCI}

Prior research has explored the potential of technology in supporting vulnerable groups of people \cite{roberson2010survival,moncur2016role,le2008designs,andalibi2016understanding, vines2013designing, vyas2017}. For example,  Massimi et al. \cite{massimi2012finding} explored the use of technology in supporting people who have experienced adverse, unpredictable, and uncontrollable events which they call ``life disruptions'' . In existing literature, we find works that deal with disruptions such as homelessness \cite{le2008designs}, death \cite{massimi2011dealing}, or domestic violence \cite{matthews2017stories, clarke2013digital}. In particular, Massimi et al. \cite{massimi2012finding} has raised three design considerations: (1) paying attention to changing social relationships, (2) considering the vulnerability of the existing infrastructure for support, and (3) upholding dignity, privacy, and safety during such times . We endorse these values but consider the term ``life disruptions'' implies a baseline with greater stability than what the trafficking survivors have. The lives of survivors have been severely disrupted from what would be considered a ``normal'' life and they have limited existing infrastructure for support. The NGO's rehabilitation program and the subsequent reintegration process are part of the journey for the survivors to achieve a ``new normal'' \cite{massimi2012finding}.  

Research work within HCI has also drawn attention to issues of social isolation experienced by vulnerable population and have presented design cases to facilitate social connections. For example, in their work with economically distressed ``women in crisis'', Capel et al. \cite{capel2017women} present a case to utilize technologies to connect women with other women in a similar situation. Dillahunt \cite{dillahunt2014fostering} argues for the need to support the building of social capital in economically distressed communities by bridging ties between different socioeconomic groups and fostering stronger ties within communities by creating identity and building trust. Similarly, other work (e.g. \cite{granovetter1977strength, putnam2000bowling, karlan2009trust}) has presented cases of informational and economic gains of social networks. Social capital arises from the embedded resources within the social network and those resources can be valuable in achieving a goal \cite{gabbay1999csc, lin2017building}. The benefits of social capital can be beyond informational and economic gains. Resnick \cite{resnick2001beyond} posits the productive resources within a social network may include a sense of collective identity, shared values, and trust. As we hear the voices of the survivors, we find that the survivors share concerns of social isolation and stigma, and have limited social capital as they move forward towards reintegration. We consider leveraging the limited but important social capital the survivors hold in further building social connections during reintegration.   

HCI research work, under the aegis of HCI for Development (HCI4D) and Information and Communication Technology for Development (ICTD), has explored opportunities for intervention in a range of social problems in developing countries (e.g. \cite{medhi2015krishipustak, vitos2017supporting}). Local culture and power relations may influence the adoption, create suspicion, or simply create indifference to intervention efforts \cite{irani2010postcolonial}. All settings are sensitive but some are more so.  Issues of power and ethics abound \cite{waycott2015ethical, munteanu2014fieldwork, redacted}. Further, post-colonial computing contends that when design space is itself being subjected to negotiation with local values, culture, and power relations, constant reflection is required on the fieldwork and on all aspects of design work including engagement with ``users'', interpretation of user needs and goals, and translation of those requirements to the designed technology \cite{irani2010postcolonial}.  

For all of these reasons, HCI research work in these areas intrinsically involves the kind of promotion of sociality that falls under the rubric of CSCW. The problems are situated in social structures and the solutions will also be social. However, arguably, these considerations mean that CSCW and HCI work in these areas must require more detailed exploration and be subject to more scrutiny than ever before settling on a direction for intervention.  Indeed, we argue that considerable analytic work needs to be undertaken before deciding whether to intervene at all, and that furthermore discussion about whether the research group has sufficient standing to intervene at all, should be part of the CSCW/HCI research corpus. 

\subsection{Broader Impact Beyond HCI}

While there has been work on the prevention of sex-trafficking in Nepal and in South Asia in general (e.g. \cite{huda2006sex, hennink2004sex}), research on trafficking survivors' experiences in rehabilitation and reintegration has been few and limited.  Some studies have involved working with survivors but few have comprehensively presented the experiences of the sex-trafficking survivors and strategies for reintegration after their return \cite{chen2003exploring}. For example, Crawford and Kaufman \cite{crawford2008sex} present findings from interviewing 20 survivors and found that 17 of them had returned to their villages after leaving the shelter homes. 11 out of those 17 were involved in some form of income-generating work. However, even though the most common skill-based training provided by anti-trafficking organizations is tailoring and sewing, only two (2) were found to be involved in similar endeavors. The majority were involved in running small grocery shops or tea shops instead, a finding corroborated by other studies \cite{sharma2015sex,poudel2009dealing}. Although the findings were encouraging for the kinds of programs most NGOs support, the paper advances concern about the lack of understanding of the factors that help in reintegration \cite{crawford2008sex}.

Similarly, in our earlier work, we present a case of four survivors who were supported by an anti-trafficking organization and were touted as successfully reintegrated. We found that, within six months, they had been forced by their families to give up their businesses \cite{redacted}. One woman was said to have been married out of the country, raising the question of whether she had been resold. We were not able to track down the actual survivors but this raised questions in our minds about the long-term prospects of survivors in current rehabilitation programs. There is merit in the programs even if all they do is protect women for a short while, but we undertook to attempt to understand the circumstances of the sex-trafficking survivors with an eye towards better long-term outcomes. Key to this is hearing their voices.

\subsection{Prior Use of Photo-elicitation and Probes}

Photography as a research method has long been used in anthropology and sociology (e.g. \cite{collier1957photography,schwartz1989visual}). Photographs can be interpreted to understand important aspects of the local culture and subsequently used for design inspiration. This is the rationale behind using cameras and photo albums in Cultural Probes \cite{gaver1999design}. Similar use of photographs has been made to get information from the participants about their situated actions through Informational Probes \cite{crabtree2003designing}. While these put the onus of interpreting photographs on the research team, photo-elicitation interviews generally use photographs taken by the participants in facilitating the interviews (e.g. \cite{le2008designs}).  

Despite capturing visible reality, photography is an inherently ambiguous medium. Ambiguity allows viewers to ``interact with the natural events depicted and draw references and significances from a broad range of events, experiences, people, and responses which they recall, derive from, relate, and attribute to the depicted contents'' \cite[pp. 39]{musello1980studying}. 

Photographs can thus be looked upon as artifacts that embody the photographers' perspective and simultaneously invite interpretation. They enable us to get a deeper understanding of the participants' perspective through the references and significances they draw. Within the CSCW community, photographs have been used in multiple ways such as in understanding practices of collaboration and sharing around photographs to inform on design of Photowares (e.g., \cite{frohlich2002requirements, crabtree2004collaborating, van2009collocated}), as a medium suited to support storytelling \cite{balabanovic2000storytelling}, and used as an  ``enjoyable activity that can help deepen personal and community relationships'' \cite[pp. 167]{frohlich2002requirements}. 

Our field has been engaging in discussions on ways to respectfully engage with people in vulnerable situations whether it be for generating design ideas or for supporting and empowering them. To that end, photo-elicitation methods have been commonly used which supports participants in choosing only those aspects of their lives that they want to share. For example, Crabtree et al. \cite{crabtree2003designing}, in their work with psychiatric patients in a hostel and elderly people at home, raise concerns about direct observation techniques. They seek to supplement such techniques with an adaptation of cultural probes and find the adaptation to be helpful both in gaining insights into user needs and in supporting user involvement early in the design process. Capel et al. \cite{capel2017women} use a self-reporting probe with videos and photographs recorded by participating women who are experiencing financial distress to generate technology design ideas. Similarly, Clarke et al. \cite{clarke2013digital} present the potential of photo-elicitation probes to support women who had experienced domestic violence and are rebuilding their lives. The participating women would bring photos, and present a video by putting the photos in sequence and adding words and sounds from audio recordings. In our social photo-elicitation method, we leverage the ambiguous, interpretative, and less-intrusive aspects of photo-elicitation. Our approach differs in that we focus on collective meaning-making rather than individual ownership of the photos and interpretations.

\section{Methodology}

This work supplements an initial ethnography with an adapted social photo-eliciting activity with survivors. The ethnography aims to understand the survivors' perspectives on their current situation and future possibilities from within their context in protected living spaces.

\subsection{Context}
After several months of email correspondence and calls to build a connection, we visited two anti-trafficking NGOs for four weeks. Both of these organizations have been operational for more than 15 years and each employs more than 100 staff members. Their work has been recognized at national and international level. The organization we call Survivor Organization (SO) was founded by survivors. Staff members at the highest-level included many who are themselves trafficking survivors. The work reported in this paper was conducted between December 2017 and January 2018 and focuses on survivors at SO. 

\subsubsection{The Partner Organization}
SO employs more than 100 people at various positions in 14 districts of Nepal. Their operations focus on three major domains: prevention of trafficking, protection, and capacity building in survivors. As part of prevention, the organization runs awareness campaigns, mobilizes local self-help groups, and provides information through counseling desks established in various districts. They also identify vulnerable girls and women and support them through skill-based training. Their work in protection involves repatriating survivors, legal support, and prosecution of those involved in trafficking, and providing psycho-social counseling. The capacity-building aspect involves providing protected living with skill-based training, and supporting survivors in income generation either through a job or starting a business.

\subsubsection{Participants}

SO's shelter homes housed 32 survivors (henceforth called sister-survivors to better reflect their own nomenclature) during the time of this study. Some of the sister-survivors worked outside the shelter homes, some worked in the organization as staff members, and others trained in the handicraft workshop that was housed within the organization's main office.

Only two of the sister-survivors had ever owned a mobile phone. No one in the shelter home was allowed to own phones. All of the sister-survivors had used landlines to make and receive phone calls. None of the sister-survivors in the shelter homes had access to the Internet. Only four of the nine sister-survivors the researcher talked with reported reading and technology skills advanced enough to have comfort reading SMS. 

\subsection{Methods}

\subsubsection{Initial Ethnography and Interviews}
The first author spent about 10 days at SO spread out over a month.  He talked with many people repeatedly during this time, including interviewing seven staff members, shadowing two key staff members at work, and conducting group discussions with staff members. The interviewed staff members included the founder, the president, a program coordinator, a media coordinator, a counselor, a legal coordinator, and a skills trainer.  He held a focus group discussion with four of the twelve sister-survivors who had recently concluded an experimental four-month-long photography program focused on angles and composition, and had published a photo-book.

Similarly, he observed sister-survivors while they were at work in the handicraft workshop which is part of their skill-building training. He was able to interact with the nine sister-survivors who were over 18 years old and did not leave the shelter home for work during the day.  All sister-survivors that participated in study were between 18 and 23 years old.

\subsubsection{Social Photo-Elicitation Method}

Substantial obstacles lay between the researcher and the sister-survivors. In addition to other aspects of his positionality, he found that staff members often answered questions that he directed to sister-survivors.  He had heard the sister-survivors' voices in some respects but he did not feel that he was hearing their voices clearly.

We had observed that there were posters adorning the interiors of the building created by the sister-survivors. Some of those posters contained photos and newspaper cutouts which suggested that the sister-survivors had worked with printed photos and so working on photos would not be very distant to the sister-survivors. Therefore, we designed an activity where we sought to leverage the ludic property inherent in photographs by adapting photo-elicitation method to mitigate the distance and hear the sister-survivors' voices in relation to their lived experience. 

The outcome would be a poster, similar to the posters created by the sister-survivors using newspaper cutouts. Although posters were familiar, we wanted to make the familiar strange in a variety of ways. The sister-survivors had never taken photos around their shelter home and their workshop. While they had used existing pictures, they had not elaborately discussed their personal interpretation of photos with each other. Moreover, we wanted to reduce the burden that sister-survivors might feel in being asked to speak as an individual voice.  Instead of asking each person about her photographs, we sought to make their contributions both \emph{social} and \emph{communal}. The blend of familiar and unconventional was intended to provoke sister-survivors to engage in discussion of personal and work-related issues, aspirations, and personal and organizational values. 

The photo-elicitation started in a fairly standard way; we gave the sister-survivors a Polaroid camera to use for two days. Before handing over the camera, we conducted a half-hour session to show them the procedure to operate a camera and asked them to operate it. They all got hands-on experience taking photos individually. Polaroids were chosen because the physical materiality of the print made it implicitly clear that the sister-survivors had the full control of the photos to the extent that they could burn them if they wanted. 

The three main innovations were that (1) instead of asking a person to present and talk about one of their own photos, the researcher-facilitator picked each photo in a random order and asked each person to talk about that photo one after another as they went around the circle; (2) at the end of each round, the group added the photo to a poster; (3) the poster was annotated with a communally agreed upon statement or comment. We conducted two sessions with five sister-survivors where each session lasted about an hour, with a week in between the sessions.

\subsubsection{Analysis Method}
Audio from the interviews and group discussions was recorded with permission from the participants. The audio was translated into English, transcribed, and coded by the first two authors. The authors independently identified concepts and themes by going over the recordings. The interpretation and meaning of the photos as expressed by the sister-survivors were used in the thematic analysis whereas the content of the photos was not used. Following this, the first two authors iterated over the codes and triangulated to identify recurring themes. Given the unique setting of the study, an inductive approach was taken during the thematic analysis. Similarly, the content of the photo-book previously created by focus group participants was coded by all the authors during a gallery walk-through session. Recurring themes were identified from the code and discussed among the authors. The categories are presented in the findings section.

\section{Findings}

The researcher found his position within the research setting more noteworthy and obtrusive than he would have liked. The personal and largely female nature of the shelter homes made him stand out. Many factors contributed to ``othering''. As a male, walking around observing the sister-survivors with a notebook seemed to make the sister-survivors self-conscious which was observed from the sister-survivors stopping their work in his presence. 

Forms of address added to the distance. Female staff members were addressed as ``mommy'' or ``older sister''. ``Sir'' was used to address all the male present including the researcher (rather than, for example, a common phrase like ``older brother''). All staff members and the researcher addressed the sister-survivors as ``younger sister'', which is, given their age, an unmarked term in Nepali. 

Furthermore, being required by the Institutional Review Board to reveal at the outset of acquaintance his relationship to a fairly well-known institution and forced to use words such as ``project'' and ``research'' added to the distance. The researcher may have been associated with donor organizations that SO depends upon. Obvious manifestations of this lay in that the researcher received inquiries about funds from some of the staff members.  Additionally, some of the sister-survivors asked him about American lifestyle and infrastructure. Their inquiries can be seen as both reflecting enough comfort to display curiosity and a heightened consciousness of him as an outsider.

The photo-elicitation method reduced the distance to some extent in that the sister-survivors were observed to be more relaxed and open. The photo-elicitation sessions were held in a familiar environment next to the handicraft workshop.  The sister-survivors and researcher sat in chairs arranged in a circle; however, since some sister-survivors got up, left, came back and moved, we felt that they knew that they were free to come and go. The sessions were conceived of as games.  They were not, for example, task-oriented. There were no right or wrong answers.  The sister-survivors were free to change the rules and in fact, did decide whether someone could ``pass'' on a photograph without talking about it. 

Most of the photographs taken by sister-survivors were of the crafts that they were learning to make. These crafts included making \textit{Pote} (Nepali bead necklaces), bracelets using glass beads, knitting of bags, scarfs, and shawls, and working on a Japanese-style of crocheted shawl called a Saori. Red and green \textit{Pote} play a role in the Nepali marriage ceremony and are important in Nepali culture as a marker of married status in women.  While the photographs focused on the craft, the discussion highlighted complexity that failed to emerge or was downplayed in our earlier interviews and group discussions.

SO's stated aims included promoting skills training to support financial independence, and in facilitating the reintegration of the sister-survivors in her own town and perhaps with her family.  While sister-survivors often endorsed both the economic and social goals and the associated values, they also expressed more nuanced aspects which at times problematized those points of view.

\subsection{Skills Training for Financial Independence}
The sister-survivors valued the craftwork not only for economic gains but for emotional value. They expressed a strong inclination to showcase their work, suggesting a sense of ownership towards the craft work. However, they also discussed the difficulty and boredom they felt in working. 

\subsubsection{Showcasing the Crafts}
\begin{figure}

\includegraphics[height=2.5in, width=4.27in]{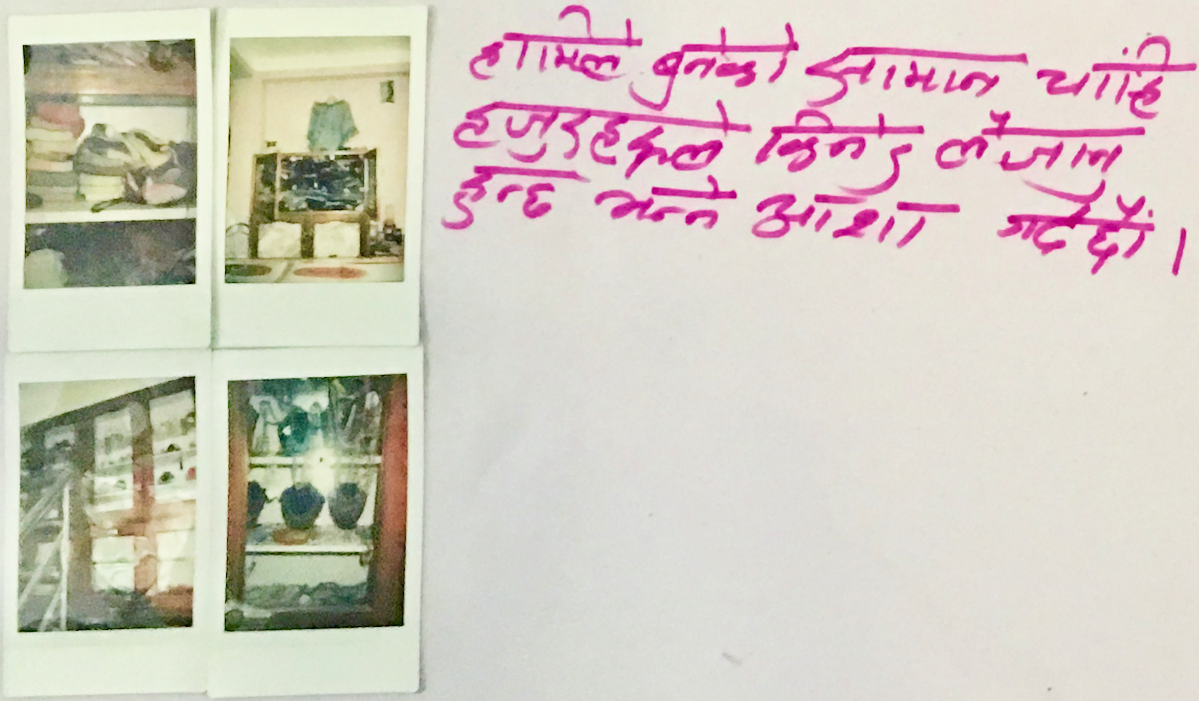}
\caption{The photo depicts a collection of handicrafts including cushion, shawls, poncho, and necklaces. Multiple participants had taken photographs of the shelf so they are grouped together. The text reads, ``We hope that you will buy the handicraft products that we have made.''}


\label{sell}
\end{figure}

Cabinets full of necklaces and various handicraft products were visible all over the building including around the staircases. These are displayed for visitors to buy, most of whom are foreigners. This appeal to visitors came up multiple times during the photo-elicitation sessions. For example, S14 said about a photo of the cabinet (Figure \ref{sell}),  ``\textit{... this is of special importance to visitors ... the main thing is, a visitor, as soon as they enter, they will see this attractive set of objects.}'' Along similar lines, S13 further added, ``\textit{... I feel like if foreign visitors come from outside, they should feel awe when they see this collection.}'' S12 further expressed her belief in the potential of selling her handicrafts outside of Nepal to places that she \textit{``can't even name.''} 

The focus on appeal was most evident in the outcome of the second session. At the beginning of the second session, a staff informed the survivors that the posters they were creating would be placed on the walls of the office building. Four out of the eight short statements written on the posters during the second session focused on trying to sell the products. In contrast, only one of the nine statements in the first session hinted at anything to do with sales.

This focus also uncovered disagreement about their perceived roles of the sister-survivors within the organization. Some wanted to write statements aimed at promoting sales. Others, like S15 below, wanted to leave advertising to others. 
\begin{quote}
S13: \textit{We should say something like `If you like this pote, please take it immediately'}

S14: \textit{Yes, that's what we should say.}

S15: \textit{No, that should be said by someone who is advertising.}

S14: \textit{Shouldn't we try to advertise?}
\end{quote}

S12 pushed the negotiation towards talking about their own activity, ``\textit{We should say that we make a lot of designs every month and wish that it could be sold quickly}'' and eventually they negotiated the sentence, \textit{``We can make different kinds of designs but you all please don't stop placing orders [for these].''} This formulation contained both references to their own activity and an appeal to visitors on behalf of the organization.

Throughout the sessions, the sister-survivors expressed a sense of ownership of the crafts and the activities. They did not express themselves as help-receivers, but rather positioned themselves as contributors to SO's activities and, like S14 and S12, sought roles as productive members of the organization. They had taken responsibility for the sales of the crafts upon themselves. 

\subsubsection{Economic Gains}

During the discussion, the value of being able to sell the handicrafts came up several times. S14 talked about the personal value of being able to make and sell \textit{Potes} to become financially independent. Similarly, S12 focused on the personal benefits of being able to make \textit{Pote}, ``\textit{... because if we learn all these skills [different handicraft], in the future, we can make these at home and sell it outside [market]. That's why I think everyone should learn Pote making}.'' S13 who preferred to work with wool and shawl-making rather than \textit{Pote} appreciated the economic benefits of her craft, ``\textit{The more yarns you make, the more profitable it is for you}.'' 

While the sister-survivors had varied opinions about the future possibilities of \textit{Pote} making to achieve economic independence, they all valued handicraft works for its economic value. S14 and S15 believed that they could ultimately be independent by creating and selling \textit{Pote}. S15 shared aspirations to sell \textit{Pote} beyond Nepal too. S12 was skeptical about \textit{Pote} but believed that she could instead sell shawls to earn money. 

However, all the sister-survivors in the photo-elicitation session expressed a negative outlook on the organization's financial gains from selling handicrafts, frequently remarking on the dwindling number of actual sales. They pointed out the lack of demand in the local market and the inability to sell as much as they had made with S14 exclaiming, ``\textit{We don't get visitors anymore. We have so many Potes outside [on the shelves], we don't know when any of them will be sold.}'' 

The fact that the sister-survivors hold contradictory views about economic gains from crafts -- that the current financial gains (to the organization) are dwindling but that they can be financially independent in the future -- raises concerns about the feasibility of relying exclusively on the limited handicraft training to promote financial independence. In the absence of other training, the sister-survivors seem to hope that what they have learned will suffice for their goals.

\subsubsection{Emotional Value of Crafts}

Along with the economic gains, the sister-survivors expressed emotional values towards the craft work. The sister-survivors mentioned their effort to not keep an idle mind. S13 while discussing the creation of shawls mentioned, ``\textit{I draw sometimes. When I go home and I am idle, thoughts creep into my mind. When I draw, the attention shifts to the drawing and takes my mind away from those thoughts.}'' 

The usefulness of handicrafts to heal from past trauma became the major focus when the sister-survivors discussed the Japanese crochet cloth-making machine. S16 informed us about the machine being used elsewhere (Japan) to argue about the value of the craft in the healing process: ``\textit{This [Saori] was helpful for the process of healing. Do you know that they have Saori machines in Japan for mentally disturbed people? So it also used here [at SO] for the process of healing.}'' In that round, the group summarized their discussion where they focused on the craft's helpfulness in healing and wrote, ``\textit{Saori is a Japanese shawl which can be taken as a form of healing. By engaging in it people can heal from different trauma}.'' 

Part of talking about the value of crafts involved imagination of possible futures. S13 mentioned that the ease of spinning yarn could provide opportunities for all family members to do something productive.  She tied this together with an image of social bonding: ``\textit{You could talk to your friends. You could be speaking while you are also engaged in work. In cold seasons, you could bask under the sun and instead of sitting idle, you could work together with friends and family to work on this}.''

Likewise, the sister-survivors talked about making the most out of the lack of resources and adjusting to it by working on their handicrafts. S14 exemplifies this initiative-taking attitude when she said, ``\textit{You don't need electricity\footnote{Power cuts are frequent in Nepal especially during the dry season. In the four winter months of 2016, Kathmandu had electricity only 8 hours/day on average.}, battery, or power to use this. You don't need help from the Internet. You don't need anything extra for this. You could relax, talk and work instead of wasting time. If you could manage your time like this, it [knitting wool] could improve your life.}"

The combination of hope and recognition of many aspects of their likely future situation that is reflected in these comments is presumably part of what the sister-survivors gain through their time in the protected living situation, but once again raises questions about longer-term pathways. At the current state, the organizations have limited resources to support on-going, long-term work on handicrafts for sister-survivors who move away from the protected living homes. As evident from earlier work \cite{crawford2008sex,redacted}, sister-survivors may be forced to forgo crafting as they become reintegrated.

\subsubsection{Difficulty and Boredom}
The discussions also included the challenges of learning the craft work. Most of them talked about how they had persevered in learning the craft. The frustration in doing some of the work was expressed such as by S14 when working on woolen shawls, \textit{``It looks easy when you see it from the outside. I found it very hard initially. I didn't know how to put the wool when I kept it on one side, the other side would be ruined.''}  She later continued, \textit{``It became irritating.''} S12 similarly expressed how hard it was for her to learn \textit{Pote} making and that she moved away from that when something else was available, never to look back, \textit{``...it [Pote] was immensely hard. I couldn't even put a thread inside the needle ... I thought shawls would be easier and I moved to work on shawls. Even now, I have not made Potes.''} It is to be noted that she persevered in learning to make shawls although she also found that hard in the beginning, ``\textit{I didn't know how to do it at first. My hands used to shiver. As time went by it became easier.}'' S13 shared her frustration and her journey moving from one craft to another not just because they were difficult but also because she did not find some interesting:

\begin{quote}
\textit{I started by learning to make necklaces but I was never interested in that. There was no shawl making back then. So, I went to learn Saori ... it was very hard. You have to put a woolen thread in one place and pull out from another and put it in another place and pull out from somewhere else! It requires a lot of effort. Once shawl making started, I left this [Saori] too.}
\end{quote}

Similar to S13, other sister-survivors also expressed boredom having to work on the same craft repeatedly. The sister-survivors had access to a limited range of craft-based activities which was limited by the availability and interest of the organization. The limitation of the range of activities came up, especially when the sister-survivors talked about the Polaroid camera. S13 remarked, ``\textit{We have knitted and made shawls but how much can one do the same work?}'' S15 further elaborated it, ``\textit{We always do the same work here and it gets boring at times. Sometimes you want something different to do or use different kinds of tools. Getting to do that [taking photos] here made me happy.}'' Her inquisitiveness and desire to use different tools led this sister-survivor to tear apart one of the Polaroid films to see what was inside it. 

There are many pragmatic reasons that SO focuses on this kind of crafting, including that the skills are attainable, the materials are inexpensive, and the products are culturally recognizable. Despite the pragmatic reasons for these crafts, these interactions and observations raised questions about whether the sister-survivors could have done something else other than the handicrafts that were available in the organization. 

In fact, earlier, SO had tried something else.  They had provided photo-editing training to sister-survivors who had produced a photo-book. The sister-survivors reported feeling overwhelmed as evident from S3's statement, ``\textit{I feel short on editing photos which we couldn't learn well. When we were learning to edit, we had never touched a laptop before to know anything [laughs] ... it would have been better if we had been able to do [editing].}'' The barrier to entry in this statement is clear. The sources of the barrier are somewhat less clear: is it engendered by the laptop itself, the editing system, something about the complexity of the editing tasks or some combination of all of these?  Did the barriers arise primarily from apprehension or from the cognitive distance of the tasks \cite{chandler1996cognitive, mayer2001cognitive}? 

Despite the negative reaction that sister-survivors reported to the editing activity, we conjecture that there may be an opportunity to build upon the sister-survivors' existing skills towards projects that they find both personally meaningful and economically beneficial. From a sufficiently abstract point of view, recent work on DIY technologies and feminist makerspaces \cite{fox2015hacking,taylor2016making, toombs2015proper} suggests extensions to the craftwork could provide an emotional and intellectual pathway towards craft and other work with more long-term economic potential.  However, moves in that direction must be taken in conjunction with conjectures about what constitutes too much of a barrier.

\subsection{Reintegrating in Society}
Like other anti-trafficking NGOs, SO's focus is on reintegration of the daughter with her family of origin.  If the family is willing and if the organization judges it safe, they engage in a process of shepherded reconciliation.  If this happens, the rationale for teaching crafts is that the ability to earn money will give the daughter more value within the family.  

However, reconciliation does not always happen. Often families reject their daughters or the situation is too dangerous, in which case, SO aims to help the women find jobs or start a business. Many staff members used the phrase ``stand on their own feet'' in describing these endeavors. They assume that the survivor's priorities would be to find a job in an office or start a business. 

Whether reunited with family or set up on their own, the current state of affairs is that the sister-survivors lose contact with one another and the NGO.  Staff responses confirm prior research suggesting that there are good reasons for this.  Being known as a sex-trafficking survivor or even as someone associated with SO can trigger stigma that threatens abuse and rejection. Yet, we observed that while the sister-survivors endorsed a vision of returning to their families with these new skills, the image of standing entirely on their own feet was more difficult.  

\subsubsection{Pain Associated with the Past}
During the focus group conducted with sister-survivors who had been involved in making the published photo-book, we asked them about their training experience. An initial task had asked the sister-survivors to write their personal stories, that is, the stories of abuse and violence that the sisters had suffered before their rescue. 

All four of the sisters who participated in the group discussion mentioned the pain they felt remembering the past events. S1 said, ``\textit{I had already forgotten it [the past events] and being reminded of it was hard. I was so sad for 2-3 days. We have left that place and moved on.}'' 

Despite the pain, all of the participants said that they completed their stories in hopes that those stories would help raise awareness.
\begin{quote}
S4: \textit{It was hard. <Redacted staff member name> asked us to write our own stories. I wrote my own story but with a pseudonym. But while I was writing the stories, having remembered incidents from the past, it hurt. I even cried. I cried not understanding why we had to revisit those old events that we had long since forgotten. At one point, I even told them that I won't write it.}

Researcher: Did you complete the story?

S4: \textit{Yes}

Researcher: What made you complete the story?

S4: \textit{To raise awareness. I felt like we should do it. If we do this, the world will see it and once they see it, they will see different ways through which people could be trafficked. I felt we should show that and so I wrote it eventually.}
\end{quote}
They valued their experiences, finding it to be of help to other people: ``\textit{People in my place may not know about these issues [trafficking]. Having gone through it, we have learned a bit and have reached here. No matter how bad it feels to us, if our stories help them, we should tell it}'' (S1).

Despite the pain, the sister-survivors expressed motivation to raise awareness about trafficking and appeared to feel a sense of agency or obligation. This seemed to us to reflect a kind of social capital.  At the current state, the sister-survivors have limited avenue to mobilize resources and lead in awareness programs, but activities based on their existing senses of purpose could enhance the sister-survivors' connection to others and pride in accomplishment.

\subsubsection{Perceived Danger in Revealing Identity}

Remaining anonymous was a high priority as highlighted in S4's statement, ``\textit{If I could tell my story without showing my face and hiding some of my fears, I would tell my story to raise awareness}'' and she elaborated further, ``\textit{If my face and voice are not recognizable, if they [public] can't figure out who I am, if they can't say that this [me] is so and so's daughter, then I would [record my story in my own voice].}'' 

\begin{figure}
\includegraphics[height=2.8in, width=3.15in]{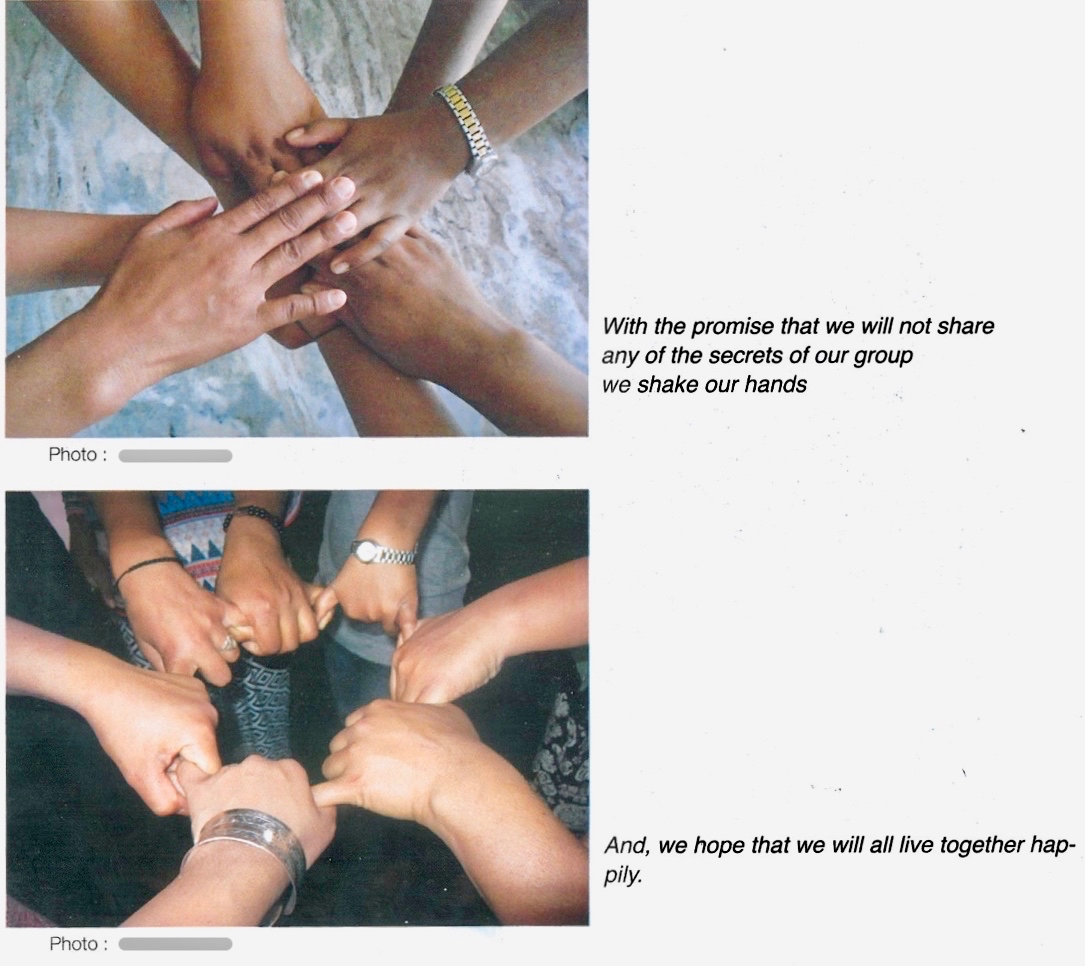}
\caption{The sister-survivors used different angles and compositional techniques to hide their identity in the photo-book.}
\label{sisterhood_book}
\end{figure}

We could see this fear of revealing identity in the photo-book as well. During the group discussion, the participating sister-survivors mentioned how they had been clever about the angle in which photos were taken such that no one in the photograph could be identified (for example Figure \ref{sisterhood_book}). 

This fear is not unfounded. Stigmatization and discrimination against sister-survivors exist in the society \cite{mahendra2001community}. A significant number of cases appear each year of sister-survivors being abused in their workplace after employers learn about their past. Staff members were aware of the issue and have made attempts to find sister-survivors jobs within trustworthy organizations. However, according to two staff members at SO, there have been ``mixed results''. 

Aligned with the principle of upholding the privacy, safety, and dignity of people who have experienced ``life disruptions'' \cite{massimi2012finding}, any design intervention in this context would have to focus on protecting the sister-survivors from being identified as trafficked persons by others in the society. Any approach, either for research or intervention, should not arouse suspicion or put the sister-survivors at risk.

\subsubsection{Desire to Belong}

During the elicitation, S15 and S16 discussed the value of working away from home and gaining membership in the outside community. S15, for example, valued the interaction with people outside of her home saying, ``\textit{Outside environment has different types of people and having a job would make it easier to be part of them. You would also build communicational and relational skills.}'' This comment is aligned with the staff's focus on independence but also projects societal connection.

\begin{figure}
\includegraphics[height=2.7in, width=4.37in]{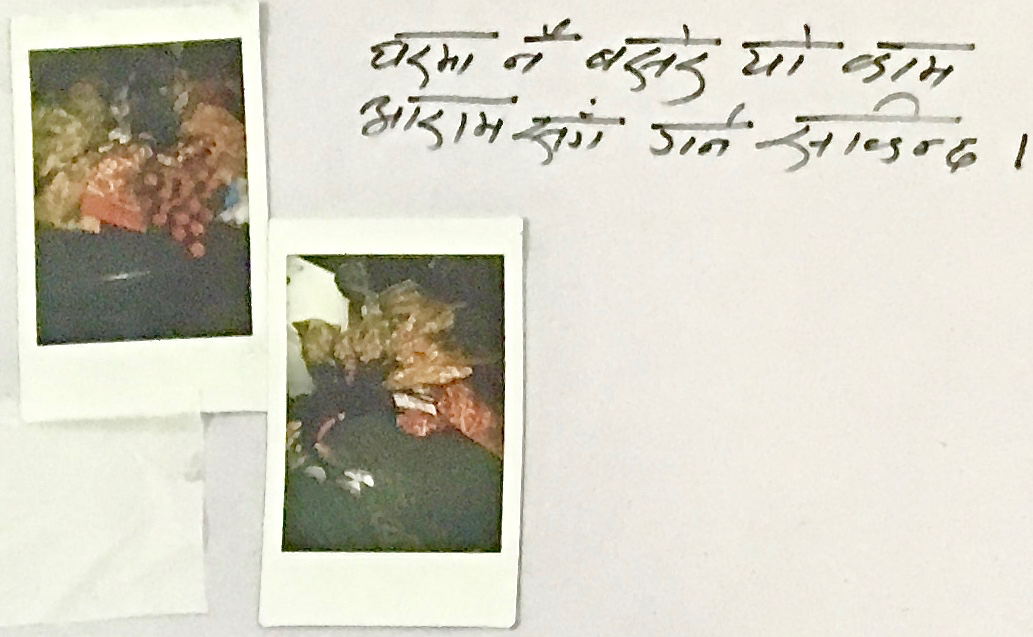}
\caption{A set of pouch made from old sari clothes and knitted decorative products created by the sister-survivors. The translated text reads ``One can easily work on these while sitting at home."}
\label{home}
\end{figure}

A different sort of wish to belong was widespread throughout the discussions: the wish to belong to a family. For instance, S13 while discussing a photo of them knitting wool (see Figure \ref{home}) mentioned, ``\textit{... you can do it [yarning wool] from home. People could also involve their brothers and sisters in the task.}'' S14 added that she prefers to work on the handicrafts ``\textit{from home. I would have children and I would drop and pick them off from school. At other times, I would work on these handicrafts.}" S12 too placed a greater value on working at home over going out, ``\textit{When you are at home, you have family members around you [laughs] I would prefer to work from home.}'' 

None of the sister-survivors said that they wanted to be simply supported by a man.  However, many appeared to want more connection than implied in simply ``going out to work.''  Belonging seemed to come from engaging with family members and other members of the community. However, many of the sister-survivors were either rejected by their families of origin or SO judged return to being too dangerous.  We did not know which sister-survivor fell into this latter category but thought it possible that these sister-survivors also shared the kind of longing for belonging and family that we heard communally. 

This raises concern given that neither organizations nor the government in Nepal have taken initiatives to support trafficking survivors to build connections with members of the community. The social network for support is limited for sister-survivors, especially for those who are rejected by their families. Research work on capital theories posits that social capital and family structure is one of the strongest factor influencing upward economic mobility \cite{dillahunt2014fostering, dominguez2003creating}. We conjecture that there may be viable opportunities to promote ``weak ties'' \cite{granovetter1977strength} between sister-survivors and members of the communities as a way to progressively build social capital. 

\subsubsection{Sisterhood and Mutual Support}
We also noted that discussion of relations between sister-survivors within the shelter homes had elements of familial environment. The photo-book created previous to our investigation and discussed by the initial focus group of sister-survivors depicted them holding each other, huddling together, and engaging in a shared activity. The narratives too described sharing personal secrets with one other, promising to keep those a secret, and a desire to live together with other sisters (see Figure \ref{sisterhood_book}). 

During the elicitation sessions, togetherness and mutual support came up multiple times. S14 mentioned how she supported newcomers by giving them scaffolds to encourage them to participate in creating \textit{Pote}. S12 was one of the recipients of S14's help who was more than happy to mention the help that S14 had provided. S13 too talked about how she seeks help from other sisters when she can't work on the \textit{Pote}, ``\textit{it's not that I don't do it at all ... I do ... when I have to work on it, I try for 4-5 times and then if I can't then I ask someone else to do it for me}.'' Similarly, when discussing a photo of working on Saori shawls, S13 talked about the help other sisters provided: ``\textit{... I went to learn Saori. It was fun in the beginning when others [sisters] would put the wool for us. Later when we had to put the wool on our own, it was very hard.}'' The shared experience involved moments of shared frustration as well.  For instance, S12 quipped how they had \textit{``thrown the shawl from the rooftop because we both couldn't do it [design the shawl].''} Support of one another was also evident during the sessions where some encouraged others to talk more during the rounds. 

Togetherness and mutual support, especially when enforced by a living situation, does not make enduring future attachment inevitable, but it does suggest that such relationships deserve further investigation as an important resource.

\section{Discussion}

This research aims to understand the circumstances and values of sex-trafficking survivors.  One goal is to develop technology that improves their situation; however, we seek to follow a path of discovery and commitment that avoid doing damage. If we conclude that the problems inherent in technological attempts to help are too risky, we will in fact find another project. We believe that working in such a sensitive situation entails the opposite of ``moving fast and breaking things'' \cite{taplin2017move}.  We need to convince ourselves as well as others that our path is deeply justified.  We consider that surfacing the considerations that would stop us from building and contemplation of the line at which we would \emph{not} intervene is an important contribution to literature and its relationship to the ethics of the endeavor \cite{baumer2011implication, vines2017our}.

\subsection{Limitations}
This discussion of sister-survivors is only one thread in the work we have already conducted. We were able to connect with staff members at SO and another organization, although only at the main offices. Other stakeholders include sister-survivors who have successfully reintegrated, donors and governmental organizations that support the NGO's and the facilities they currently use to market the products of the sister-survivors' crafts outside the organization itself.  These efforts sit between efforts to prevent trafficking in the first place and to improve the position of all women in Nepal. 

Furthermore, we do not claim that our four-week study makes us experts on this setting. But we learned a great deal about the general situation of sex-trafficking survivors and we feel that we were able to hear some important nuanced elements of what mattered to the sister-survivors at SO. 

\subsection{Social Photo-Elicitation}

We drew upon well-known features of photo-elicitation.  The initial ability to use the camera whenever the sister-survivors wanted was important to 1) make them comfortable with the tool and the method and 2) win their trust by handing over a relatively expensive camera to them. 

We considered it important to begin to hear the voices of the sister-survivors but we did not want to cause discomfort. To make the sister-survivors more comfortable, we adapted the methods to emphasize the social and communal. After hearing about the difficulty involved in making stories about their experiences as trafficked persons, we deliberately refrained from probing that pain.  We similarly refrained from such sensitive topics as whether any particular woman was likely to be accepted back by her family or not.  Instead, we tried to create a ludic experience.  

Photographs and their descriptions were not treated as the personal possessions of the photographer but rather as elements shared in a lighthearted group activity.  By asking each participant to comment on each photograph, we made the activity playful and social. In contrast to photo-elicitation studies (e.g. \cite{clarke2013digital,capel2017women,le2008designs}), we diminished individual ownership of the photographs and the degree to which each person was ``on the hook'' to create an independent interpretation. This tactic, which might not work in all settings, was appropriate in a setting where participants were particularly likely to shy away from standing out as individuals \footnote{ Nepal can be characterized as, in general, being a more communal society than the United States, but sister-survivors might be more reluctant than most to assert themselves as individual contributors.}. The open-ended subjectivity of the photographs was used as a resource to facilitate a deeper engagement in discussions. The discussions were not limited to the significance depicted in the image but rather what it meant to each of the participants. It drew out pluralistic voices because each had different experiences of the same setting. By asking them to come up with a single descriptive sentence or phrase, we also increased the degree to which they were creating a communal product.

Taking our cues from the sister-survivors, we wanted the discussion to remain in their comfort zones. Laughter, participation, and freedom of movement suggest that we succeeded to some extent. The outcome began to give us a sense of what the sister-survivors valued.

\subsection{What We Heard}

We gained nuanced messages from the photo-elicitation experiences.  First, the sister-survivors' relationship to crafting was complex.  On one hand, they wanted to showcase their crafts, thought that crafting was therapeutic, had persevered through difficulty to learn the craft work, and could imagine futures involving crafting.  On the other, they worried about whether the crafts had financial value, and found crafting sometimes boring.  We wondered whether other crafting possibilities might be useful.

Their ideas about reintegration were also complex.  While none expressed the desire to simply be supported by family or husband, they envisioned work in a family context.  Because such a context seemed by-and-large unlikely, we noted that a related potential resource for social connection lay in the relationships they formed to other sister-survivors. We noted the lack of opportunities for sister-survivors to engage with other members in the community through meaningful activities during their reintegration phase.

\subsection{Designing to Support Building Social Capital}

Right now, sister-survivors leave the protected living situation to a mostly unknown future.  We have identified elements of strength in their situations.   There may be other elements that we have not discovered, but our future work will build on elements of strength that we find.  Two resources that we see now could be built upon to further their lives after leaving the protected living situation.  The first is their craft and the second is their ties to one another. One major problem plagues design in both directions.

\subsubsection{Building on their Craft}

The ultimate outcome wanted by the sister-survivors is to be reintegrated in society with dignity, with a sense of belonging. For greater acceptability in society, the survivors would have to be in situations that allowed them to create ties with other members of the community in which they would be treated as equals. To that end, their sense of agency, ownership, and skills in making crafts could be leveraged as resources. 

In earlier works, \cite{roberson2010survival}, based on their study of homelessness, suggest finding commonalities among different socioeconomic groups which could then be used in building ties. Similarly, Dillahunt \cite{dillahunt2014fostering} advocated the need to increase sharing of common physical spaces across diverse socioeconomic groups and thereby ``link'' \cite{woolcock1998social} people with low social capital with those with more capital. While the research works highlight economic gains, the increased interactions where sister-survivors are shown to be as productive members of the community as anybody else would help towards greater social acceptability. 

Many women in Nepal could profit from crafting skills, especially if those skills were constructed in such a way as to lead to growing capability.  A design sensibility is to leverage the sister-survivors' existing skills in group settings for mutual growth and profit while working towards shared goals. Such settings might encourage interactions between the sister-survivors and members of the community.

\subsubsection{Sustaining Existing Connections}
Sister-survivors appear to rely on their relationships with one other in the protected living environment. This constitutes social capital that could be beneficial for long-term emotional support and collective action \cite{resnick2001beyond}. Right now, there is no support for long-term connection. 

Other research work on fostering connections and support has relied on access and use of technology \cite{andalibi2016understanding,roberson2010survival, moncur2016role}. Similar to Massimi et al.'s suggestion \cite{massimi2012finding}, one design sensibility would be to provide a way to sustain the connection. 

In the context of the sister survivor's, barriers to technology use are formidable. The sister survivors lack access to smartphones and web technologies, and are, by-and-large, semi-literate. While a potential solution could be working on voice-based systems to build a community that is accessed through non-smart phones (e.g. \cite{vashistha2015sangeet, raza2013job}) or public kiosks (e.g. \cite{plauche2007speech}), mediating interactions that require explicit actions (such as to make a phone call) may be problematic. 

Another avenue would be to explore the design of technological tools that are present in the background and support a feeling of intimacy and togetherness such as Messaging Kettle \cite{brereton2015messaging} or Ambient Birdhouses \cite{brereton2017ambient} to allow sister-survivors to feel together without being brought to the attention of others in society.

\subsubsection{Supporting Obfuscation}

Both of these design directions face a common, formidable challenge.  The sister-survivors were well aware of the issue of stigma. Whether in their families of origin or relocated to a new village, they risk shunning or violence if they are known to have been trafficked or even to have been associated with SO. The sister-survivors' desire to hide their identity suggests that any design intervention would have to be able to be used imperceptibly and any artifacts would have to be deniable and arouse no questions or suspicion.  This is a particularly difficult design task when we anticipate that the sister-survivors will have virtually no privacy after release.

While theorists and practitioner advocate building trust to foster social capital \cite{resnick2001beyond,dillahunt2014fostering,gabbay1999csc}, we see a caveat to the principle here. Hiding aspects of the past events may be necessary to build social capital in this context, and whether necessary or not, the sister-survivors must be in control and comfortable with any risk. The case of sex-trafficking survivors in Nepal raises the issue of designing systems that provide cover and deniability to a new level. 

One design approach is to embrace obfuscation as a design principle.  The need is analogous to the need for privacy under large-scale surveillance context. Brunton and Nissenbaum make a case for obfuscation as a design response to protect against such surveillance where the users have little agency to opt-out and there are no facilities to support privacy \cite{brunton2015obfuscation}. They further argue that ``obfuscation offers a means of striving for balance defensible when it functions to resist domination of the weaker by the stronger'' \cite[pp. 83]{brunton2015obfuscation}. Other work has presented case studies where such deception are considered as a form of empowerment \cite{van2016computationally}. Moreover, research suggests that interactions where a person's identity is not revealed may allow people to overcome the perceived stereotypes or take productive actions \cite{resnick2001beyond}. While ethical concerns are justified in comfort with such deception and dishonesty \cite{van2016computationally,brunton2015obfuscation}, the argument could be that the risk attached to being identified as a trafficked person outweighs the concern. Following Rawl's Maximin principle \cite{rawls1974some,brunton2015obfuscation} of maximizing gains for the minimum (underprivileged), we contemplate the design principle of supporting obfuscation in the case at hand. For example, we might be able to set up social groups which treat sex-trafficking survivors amongst many others, as valued for their crafts.

Ideally, we would want Nepali society to not stigmatize and discriminate against the sister-survivors. However, such a change in public attitude takes time and may not be attainable through design interventions. Until such changes occur, many interventions would have to integrate support for obfuscation that provides cover and facilitate deniability to the sister-survivors.

\subsection{Future Work}

In contemplating action, first, we are aware that we have only scratched the surface of understanding the sister-survivors and their situation.  We need to know more. Our commitment to a better future informs us to refrain from design intervention that specifically targets sister-survivors which could cause suspicion and potentially pose a risk to their reintegration process and even their lives. At the current moment, we do not have a strong strategy to fight against social stigma in society more generally nor do we have a method of protecting individual sister-survivors from the stigma that might fall upon them if the designed intervention fails to protect them.  Considering the potential risk that our design intervention could bring, a viable option is to \emph{not} intervene. 

However, inaction is difficult to accept, having observed their limited future opportunities and questionable long-term outcomes. We are cautiously moving forward with a design direction presented here: creating a common space for \emph{all} women in the community to come together and work on crafts. We will promote group activity by providing lessons in marketing products using publicly-placed computers and the Internet.  These lesson plans have two-fold aims: 1) broaden the notion of future economic possibilities of craft work for the sister-survivors, and 2) bring to the forefront the common role of everyone as ``students'' and use that as an avenue for socialization and further interactions.  We consider the approach as a first step towards a long-term engagement to further our understanding of the situation of the sister-survivors during reintegration. 

\section{Conclusion}
\label{con}
We created a social photo-elicitation activity as part of trying to understand the situation of sex-trafficking survivors in Nepal. The activity positioned the intentions of the authors -- especially those from Nepal -- in a ``proto-reflective'' situation; that is, taking action with an intervention will perturb the system, but inaction is problematic. We learned about their complex relationships with crafting and their envisioned reintegration with the society at large.

A first read of the findings from this work might suggest that there is no design space for responsible technological intervention with this community in this setting. In fact, a closer read of the findings along with a broader interpretation of what CSCW research is about supports the further investigation of the particulars of the sister-survivors' longer-term needs and their relationship to social technologies. That is, they are embedded in a social network of great complexity that is in turn embedded in a highly structured, but fraught culture; the challenge now is to configure technological social infrastructure in that context.

Our findings obey the CSCW and HCI mandates of encountering people in the complexities of their lived lives. The complexities of these lived lives deepen our understanding of the limited but important social capital that the survivors hold.  Beyond the particulars of this situation, it is crucial that CSCW continue to grow in its understanding of ethical responsibilities and its contemplation of what elements of a situation or a design allow ethical intervention.  The two key questions in design are, ``How do you know what is the right thing to do?'' and ``How do you know when you have done it?''

\bibliographystyle{ACM-Reference-Format}
\bibliography{sample-bibliography}

\end{document}